\documentclass[12pt]{iopart}
\usepackage{cite}

\newcommand{\mean}[1]{\left\langle #1 \right\rangle}
\newcommand{\smean}[1]{\langle #1 \rangle}
\newcommand{\Nev}{N_{\rm evts}}

\begin{document}

\title{Anisotropic flow from Lee--Yang zeroes: a practical guide}

\author{\underline{N Borghini}\dag, R S Bhalerao\ddag\  and J-Y Ollitrault\dag}

\address{\dag\ Service de Physique Th\'eorique, CEA-Saclay, 
  F-91191 Gif-sur-Yvette cedex, France}

\address{\ddag\ Department of Theoretical Physics, 
  Tata Institute of Fundamental Research, Homi Bhabha Road, Colaba, 
  Mumbai 400 005, India}

\begin{abstract}
We present a new method to analyze anisotropic flow from the genuine 
correlation among a large number of particles, focusing on the practical 
implementation of the method. 
\end{abstract}



\section{Introduction}
\label{s:intro}

Anisotropic collective flow is defined as a correlation between the azimuthal
angle $\phi$ of an arbitrary particle and the azimuth $\Phi_R$ of the impact 
parameter in a non-central nucleus--nucleus collision. 
It is commonly characterized by the Fourier coefficients of the single-particle 
distribution~\cite{Voloshin:1994mz}
\begin{equation}
\label{vn}
v_n \equiv \mean{\cos(n(\phi-\Phi_R))}, 
\end{equation}
where $\smean{\ldots}$ here denotes an average over many particles and events. 
In particular, elliptic flow $v_2$ is recognized as a sensitive probe of 
thermalization at RHIC~\cite{Gyulassy:2004}. 

While anisotropic flow is by definition a collective effect that involves many 
particles, it has mostly been analyzed using methods based either on a study 
of two-particle correlations~\cite{Danielewicz:hn,Ollitrault:1997di,%
  Poskanzer:1998yz} or on the cumulants of correlations between a few (in 
practice, up to 8) particles~\cite{Borghini:2001vi}.
We recently proposed a new method of analysis~\cite{Bhalerao:2003yq,%
  Bhalerao:2003xf} that remedies this limitation, and extracts flow from the 
correlation between a large number of particles instead of only a few. 
In the following, we introduce the practical recipes for implementing the 
method, referring the reader to Refs.~\cite{Bhalerao:2003yq,Bhalerao:2003xf} 
for theoretical justifications.

\section{Integrated flow}
\label{s:integrated}

For a given centrality bin, the first step in the analysis is to obtain an 
estimate of the flow integrated over some phase-space region (typically 
corresponding to the acceptance of a detector or a set of detectors). 
We define {\em integrated flow\/} as the average over events
\begin{equation}
\label{Vn}
V_n \equiv \mean{\sum_{j=1}^M w_j \cos(n(\phi_j-\Phi_R))},
\end{equation}
where the sum runs over all particles detected in an event. 
As is usual in flow analyses, $w_j$ is a weight attributed to the $j$-th 
particle so as to maximize the integrated flow value. 
Further details regarding the choice of weights can be found in 
Ref.~\cite{Borghini:2001vi}. 
With unit weights, and neglecting for simplicity multiplicity fluctuations, 
$V_n = Mv_n$, where $v_n$ is an average of the Fourier coefficient~(\ref{vn}) 
over the phase space covered by the detector.

To derive an estimate of the integrated flow $V_n$ (for practical purposes, 
$n=2$ at RHIC and LHC, $n=1$ at AGS and below), one should first compute for 
each event the complex-valued function\footnote{
  In Refs.~\cite{Bhalerao:2003yq,Bhalerao:2003xf} we used a different 
  generating function.
  The one defined in Eq.~(\ref{g(ir)}), which is introduced in Appendix~A of
  Ref.~\cite{Bhalerao:2003xf}, actually yields more accurate results for higher
  harmonics $v_{2n}$, $v_{3n}$ when analyzing differential flow. 
  However, both forms of generating functions are actually equivalent for the 
  most part, and the computations of Ref.~\cite{Bhalerao:2003xf} can easily be 
  adapted to the present function Eq.~(\ref{g(ir)}).}
\begin{equation}
\label{g(ir)}
g^\theta(\rmi r) \equiv 
\prod_{j=1}^M [ 1 + \rmi r\,w_j\cos(n(\phi_j-\theta))]
\end{equation}
for various values of the real positive variable $r$ and of the angle $\theta$ 
($0\leq\theta <\pi/n$; in practice, 4 or 5 equally spaced values of $\theta$ 
are enough as shown in Ref.~\cite{Bhalerao:2003xf}). 
The $\phi_j$ are the {\em measured\/} azimuthal angles of the particles, 
using a fixed reference in the laboratory, and the product runs over {\em all\/} 
particles.
Please note that the method is stable against effects, like multiple hits or 
showering, which bias the results of other methods of analysis, so that one 
should not refrain from using all detected particles, combining information 
{}from different detectors: increasing the multiplicity results in smaller 
statistical uncertainties on the flow estimates. 

Together with values of $g^\theta(\rmi r)$, one should in each event compute 
the sums 
\begin{equation}
\label{Qx,Qy}
Q_x \equiv \sum_{j=1}^M w_j\cos(n\phi_j), \qquad
Q_y \equiv \sum_{j=1}^M w_j\sin(n\phi_j), 
\end{equation}
as well as their squares ${Q_x}^2$ and ${Q_y}^2$. 

Next, one should average $g^\theta(\rmi r)$ over events for each value of 
$r$ and $\theta$: 
\begin{equation}
\label{G(ir)}
G^\theta(\rmi r)\equiv \mean{g^\theta(\rmi r)}_{\!\rm evts} \equiv
\frac{1}{\Nev} \sum_{\rm events} g^\theta(\rmi r),
\end{equation}
where $\Nev$ is the number of events used in the analysis.
This is also a good time to compute the averages over events $\smean{Q_x}$, 
$\smean{Q_y}$ and $\smean{{Q_x}^2+{Q_y}^2}$. 

For every $\theta$ value, one must then look for the position $r_0^\theta$ of 
the first positive minimum of {\em the modulus\/} $|G^\theta(\rmi r)|$. 
An estimate of the integrated flow $V_n$ is given by
\begin{equation}
\label{Vn(theta)}
V_n^\theta\{\infty\} \equiv \frac{j_{01}}{r_0^\theta},
\end{equation}
where $j_{01} \simeq 2.40483$ is the first zero of the Bessel function $J_0$. 
If the detector acceptance has reasonable azimuthal symmetry, the estimates do 
not depend on $\theta$ up to statistical fluctuations (see below).\footnote{
  For an anisotropic detector, $V_n^\theta\{\infty\}$ shows an oscillatory 
  pattern that can be computed in terms of the Fourier coefficients of the 
  detector acceptance-efficiency profile.
  There is an extra proportionality factor between $V_n\{\infty\}$ and $V_n$ 
  that can also be calculated and is close to unity in most cases, see 
  Ref.~\cite{Bhalerao:2003xf} for details.}
One eventually averages $V_n^\theta\{\infty\}$ over $\theta$.
This yields a new estimate $V_n\{\infty\}$ with smaller statistical errors.
This estimate is then used to compute the resolution parameter, which measures 
the relative strength of flow compared to finite-multiplicity fluctuations: 
$\chi\equiv V_n\{\infty\}/\sigma$~\cite{Ollitrault:1997di}, with $\sigma$ 
given by
\begin{equation}
\label{sigma}
\sigma^2 \equiv 
\mean{{Q_x}^2 + {Q_y}^2} - \mean{Q_x}^2 - \mean{Q_y}^2 -V_n\{\infty\}^2. 
\end{equation}

The relative statistical error on $V_n^\theta\{\infty\}$ due to the finite 
number of events is
\begin{equation}
\label{dVn(theta)}
\fl\frac{\mean{(\delta V_n^\theta\{\infty\})^2}}{V_n^2}=
\frac{1}{2\Nev\, j_{01}^2\,J_1(j_{01})^2} \left[
  \exp\!\left(\frac{j_{01}^2}{2\chi^2}\right) +
  \exp\!\left(\!-\frac{j_{01}^2}{2\chi^2}\right) J_0(2j_{01}) \right],
\end{equation}
where $J_1$ is the spherical Bessel function of order 1. 
The statistical uncertainty on the estimate $V_n\{\infty\}$ is a about a 
factor of 2 smaller~\cite{Bhalerao:2003xf}.

\section{Differential flow}
\label{s:differential}

Once integrated flow estimates have been obtained, one can turn to the analysis
of {\em differential flow}, i.e., the flow of particles of a given type in a 
definite phase-space window, which we shall call ``protons'' for the sake of 
brevity. 
A ``proton'' azimuth will be denoted by $\psi$, and the corresponding 
differential flow $v_p(p_T,y)$ by $v'_p$. 
Using an estimate of integrated flow in harmonic $n$, as e.g. 
$V_n^\theta\{\infty\}$, one can derive an estimate of $v'_p$ in any harmonic 
$p$ which is a multiple of $n$, i.e., $p=mn$ with $m$ integer. 

Now, for a given angle $\theta$, with the help of the values of $r_0^\theta$ 
and $V_n^\theta\{\infty\}$ determined following the recipe of 
Sec.~\ref{s:integrated}, an estimate of $v'_{mn}$ is given by
\begin{equation}
\label{vmn(theta)}
\fl
\frac{{v'}_{mn}^\theta\{\infty\}}{V_n^\theta\{\infty\}} \equiv 
\frac{J_1(j_{01})}{J_m(j_{01})}\, {\rm Re}\left(
  \frac{\displaystyle\mean{g^\theta(\rmi r_0^\theta)\,
      \frac{\cos(mn(\psi-\theta))}
      {1+\rmi r_0^\theta w_\psi\cos(n(\psi-\theta))}}_{\!\!\psi}}
    {\displaystyle\rmi ^{m-1}
      \mean{g^\theta(\rmi r_0^\theta)\sum_j  \frac{w_j\cos(n(\phi_j-\theta))}
        {1+\rmi r_0^\theta w_j\cos(n(\phi_j-\theta))} }_{\!\!\rm evts}}  \right).
\end{equation}
In the denominator, the average is over events and the sum runs over all 
particles in each event. 
By contrast, the average $\smean{\ldots}_\psi$ in the numerator is over 
{\em protons}, and we have denoted by $w_\psi$ the weight associated with a
proton. 
Please note that the sum in the denominator need only be computed once per 
event (it is actually the derivative of $g^\theta(ir)$ at the minimum 
$r_0^\theta$), while the quantity to be averaged in the numerator varies from 
one proton to the other, even for protons within the same event. 
Finally, Re denotes the real part of the (complex-valued) ratio. 

Denoting by $N'$ the total number of ``protons'' in the phase-space bin under
study, the statistical uncertainty on the estimate ${v'}_{mn}^\theta\{\infty\}$ 
is
\begin{equation}
\label{dv'mn(theta)}
\fl
\mean{(\delta{v'}_{mn}^\theta\{\infty\})^2} = 
\frac{1}{4N'J_m(j_{01})^2} \left[ 
  \exp\!\left(\frac{j_{01}^2}{2\chi^2}\right) +
  (-1)^m\exp\!\left(\!-\frac{j_{01}^2}{2\chi^2}\right) J_0(2j_{01}) \right].
\end{equation}
As in the case of integrated flow, averaging the various estimates
${v'}_{mn}^\theta\{\infty\}$ results in a new estimate $v'_{mn}\{\infty\}$ 
with reduced statistical error bars (by a factor $\simeq$ 2). 
Regarding the effects of detector anisotropies, they are the same as above 
too~\cite{Bhalerao:2003xf}: 
a $\theta$-dependence of ${v'}_{mn}^\theta\{\infty\}$ and a multiplicative 
factor between the ``true'' $v'_{mn}$ and its estimate $v'_{mn}\{\infty\}$.

Finally, let us briefly mention systematic errors inherent to the method. 
A careful study in Ref.~\cite{Bhalerao:2003xf} allowed us to conclude that the 
relative error due the interplay of nonflow effects and higher harmonics (in 
particular $v_{2n}$) is of order
\begin{equation}
\label{dVn(inf)syst}
\frac{\delta v'_{mn}\{\infty\}}{v'_{mn}} = \Or\!\left(\frac{1}{M}\right) + 
\Or\!\left(\frac{1}{(Mv_n)^2}\right) + \Or\!\left(\frac{v_{2n}}{Mv_n^2}\right).
\end{equation}
The relative error on integrated flow estimate $V_n\{\infty\}$ is of the same 
order of magnitude. 
Actually, in the analysis of higher harmonics ($m>1$), an extra error term 
arises
\begin{equation}
\delta v'_{mn}\{\infty\}= \Or\!\left(\frac{v'_{(m-1)n}}{M v_n}\right).
\end{equation}
Unlike the previous term, this is an absolute systematic error, not a relative 
error on the flow.
With any other method of flow analysis, the systematic error will always be 
larger (or at least equal).

\section{Discussion}

The method of analysis we presented above is simple to implement: compute 
values of the generating function~(\ref{G(ir)}) and find the first minimum of 
its modulus, then Eq.~(\ref{Vn(theta)}) gives you the integrated flow. 
Knowing the position of the minimum, you can then compute in a second pass 
through the data the quantities in the right-hand side of Eq.~(\ref{vmn(theta)}) 
and deduce differential flow. 
Nothing more is required. 

In addition, the method is conceptually rich. 
The minimum of $|G^\theta(\rmi r)|$ is in fact compatible with a zero of 
$G^\theta(z)$, where $z$ is a complex variable. 
The mere existence of such a zero {\em close to the origin and scaling with
  the inverse of the system size} ($V_n$ in Eq.~(\ref{Vn}) is roughly 
proportional to the multiplicity $M$) signals the presence of collective 
effects.  
This is analogous to Lee--Yang theory of phase transitions~\cite{LeeYang:1952} 
in which the zeroes of the partition function come closer to the origin with 
increasing system size if, and only if, there is a phase transition. 
Since anisotropic flow is a collective effect, Lee--Yang zeroes definitely are 
the most natural method to analyze flow.

\end{document}